\documentclass[10pt,twocolumn,letterpaper]{article}

\usepackage[pagenumbers]{wacv} % To force page numbers, e.g. for an arXiv version

\usepackage[dvipsnames]{xcolor}
\usepackage{algorithm}
\usepackage{algpseudocode}
\usepackage{graphicx}
\usepackage{amsmath}
\usepackage{amssymb}
\usepackage{booktabs}
\usepackage[dvipsnames]{xcolor} 
\usepackage{xspace}
\usepackage[normalem]{ulem}
\makeatletter
\DeclareRobustCommand\onedot{\futurelet\@let@token\@onedot}
\def\@onedot{\ifx\@let@token.\else.\null\fi\xspace}
\def\eg{\emph{e.g}\onedot} 

\def\ie{\emph{i.e}\onedot}

\def\etc{\emph{etc}\onedot} 

\def\wrt{w.r.t\onedot} 

\def\etal{\emph{et al}\onedot}
\makeatother

\usepackage[pagebackref,breaklinks,colorlinks]{hyperref}

\usepackage[capitalize]{cleveref}
\crefname{section}{Sec.}{Secs.}
\Crefname{section}{Section}{Sections}
\Crefname{table}{Table}{Tables}
\crefname{table}{Tab.}{Tabs.}

\def\wacvPaperID{187} % *** Enter the WACV Paper ID here
\def\confName{WACV}
\def\confYear{2025}

\begin{document}

\title{SeCo-INR: Semantically Conditioned Implicit Neural Representations for Improved Medical Image Super-Resolution}

\author{Mevan Ekanayake, Zhifeng Chen, Gary Egan, Mehrtash Harandi, Zhaolin Chen\\
Monash University\\
Melbourne, Australia\\
{\tt\small \{mevan.ekanayake, zhifeng.chen, gary.egan, mehrtash.harandi, zhaolin.chen\}@monash.edu}
}

\maketitle

%%%%%%%%% ABSTRACT
\begin{abstract}
Implicit Neural Representations (INRs) have recently advanced the field of deep learning due to their ability to learn continuous representations of signals without the need for large training datasets. Although INR methods have been studied for medical image super-resolution, their adaptability to localized priors in medical images has not been extensively explored. Medical images contain rich anatomical divisions that could provide valuable local prior information to enhance the accuracy and robustness of INRs. In this work, we propose a novel framework, referred to as the Semantically Conditioned INR (SeCo-INR), that conditions an INR using local priors from a medical image, enabling accurate model fitting and interpolation capabilities to achieve super-resolution. Our framework learns a continuous representation of the semantic segmentation features of a medical image and utilizes it to derive the optimal INR for each semantic region of the image. We tested our framework using several medical imaging modalities and achieved higher quantitative scores and more realistic super-resolution outputs compared to state-of-the-art methods.

\end{abstract}

%%%%%%%%% BODY TEXT

\section{Introduction}

Advancements in deep learning (DL) research have revolutionized medical image super-resolution~\cite{zhao2019applications,zhao2020smore,qiu2020super, chen2023deep}. These DL methods often rely on supervised learning techniques to transform low-resolution images into high-resolution ones, using large cohorts of datasets. While this approach demonstrates remarkable success in specific scenarios, its performance deteriorates when faced with out-of-distribution cases, as many models struggle to generalize beyond the extent of the training data. Furthermore, for medical imaging applications, it is challenging to obtain massive amounts of training data due to privacy concerns and high acquisition costs. As a result, exploring unsupervised or semi-supervised DL methods for image super-resolution has become important with the ultimate goal of advancing the precision and reliability in medical imaging~\cite{gundogdu2024self}.

Implicit Neural Representations (INRs) have recently advanced the field of signal representation due to their ability to learn continuous representations of signals using a discrete set of samples without the requirement of large training datasets~\cite{sitzmann2020implicit}. In contrast to conventional approaches that discretely store signal values on coordinate grids, INRs train neural networks, particularly Multilayer Perceptrons (MLPs), to represent continuous signals~\cite{mildenhall2021nerf}. The objective is to approximate the intricate connection between coordinates and their corresponding signal values, ultimately producing a continuous representation of the signal. Due to their immense success in signal representation and solving inverse problems, INRs have been utilized in many tasks recently~\cite{chen2021learning,molaei2023implicit,dupont2021coin}. Some of the recent medical imaging-specific advancements include arbitrary scale 3D super-resolution~\cite{wu2022arbitrary,wu2021irem}, assessing tumor progression with sparsely sampled images~\cite{shen2022nerp}, jointly learning different medical imaging modalities of complementary views~\cite{mcginnis2023single} and reconstruction of isotropic images from anisotropic images~\cite{zhang2023self} belonging to a variety of medical imaging modalities like magnetic resonance imaging (MRI) and computed tomography (CT) imaging.

While INR techniques have been recently explored to enhance medical image super-resolution~\cite{mcginnis2023single,ye2023super}, their potential to benefit from localized anatomical priors has not been thoroughly investigated. Medical imaging modalities contain valuable local priors that could effectively enhance the representation of images~\cite{danielyan2011bm3d,oszust2020no}. For instance, different cortical regions of a brain image exhibit various contrasts and geometrically varied anatomical structures due to the unique signal attributes captured by the imaging system~\cite{mai2015atlas,evans2012brain}. In this work, we hypothesize that this semantic contrast and shape information can benefit INRs in learning the continuous representation of a medical image. To this end, we proposed a framework, referred to as the Semantically Conditioned INR (SeCo-INR), which learns a continuous representation of the semantic segmentation features of an image and utilizes it to derive the optimal INR parameters for each semantic region of the image through a data-driven process. Through comprehensive testing on various medical imaging datasets, our framework demonstrated higher quantitative scores and produced realistic super-resolution outputs, outperforming the state-of-the-art methods.

\begin{figure*}[t]
\centering
\includegraphics[width=0.98\textwidth]{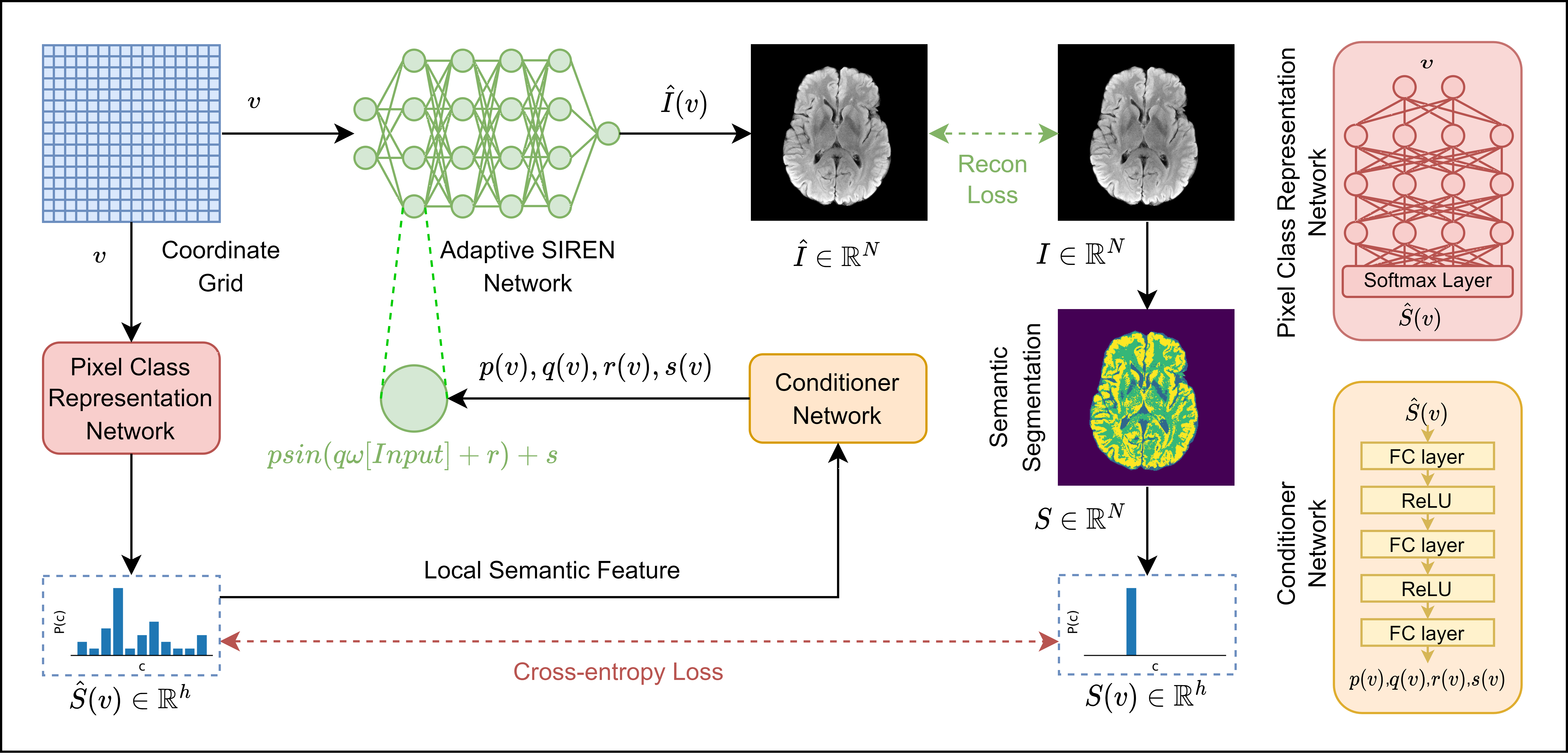}
\caption{Illustration of SeCo-INR: Semantically Conditioned Implicit Neural Representations framework for Medical Image Super-Resolution. The Pixel Class Representation Network learns a continuous representation of the semantic segmentation mask, which provides conditioning to the Adapting SIREN Network via the Conditioner Network. This process enables the learning of a more accurate implicit neural representation of the underlying medical image, allowing robust super-resolution capabilities.}
\label{fig_framework}
\end{figure*}

\section{Methods}

\subsection{Implicit Neural Representations}

Implicit Neural Representations encode an underlying continuous signal, $G$ via a deep neural network, $f_{\theta}:\mathbb{R}^{M} \rightarrow \mathbb{R}^{P}$, where $M$ is the dimension of input coordinates and $P$ is the dimension of signal value at that particular coordinate. As the deep neural network, often an MLP with 3-4 layers is utilized with suitable activation functions in between. This network learns a functional mapping between the coordinates, $v\in\mathbb{R}^{M}$ (\eg, $v=(x,y)$ for 2D, $v=(x,y,z)$ for 3D, \etc) and the underlying signal, $G(v)\in \mathbb{R}^{P}$ (\eg, intensity, color, occupancy in 3D space, camera view, \etc)~\cite{xie2023diner}. This can be achieved by optimizing the discrepancy between a set of discrete samples from $G(v)$ and the corresponding network output $f_\theta(v)$, given by $\| f_\theta(v) - G(v) \|_2^2$. Each layer of $f_\theta$ is activated using $y_l=\sigma(W_l y_{l-1} + b_l),$ where $l=1, 2, \ldots, L-1$, and $L$ is the total number of layers in the MLP. Here $W_l\in \mathbb{R}^{d_{l-1} \times d_l}$ and $b_l \in \mathbb{R}^{d_l}$ denote the weights matrix and the bias vector at the $l^\text{th}$ layer of the network. $y_l\in \mathbb{R}^{d_l}$ is the output of the $l^\text{th}$ layer and $\sigma$ is the activation function.

When Rectified Linear Unit (ReLU) is used as the activation function, \ie, $\sigma(x)=\max(0, x)$, the network tends to favor the representation of low-frequency signals which often leads to lower-quality signal representations~\cite{saragadam2023wire}. To address this issue, Sitzmann \etal~\cite{sitzmann2020implicit} proposed sinusoidal representation networks (SIRENs), which utilized periodic activation functions to fit the signal as well as its higher-order derivatives, \ie, $\sigma(x)=\sin(\omega_0 x)$. Although SIREN outperforms ReLU in terms of image representation and various other tasks, there remains room for improvement in managing the sine activation function more effectively. Subsequent works demonstrated that the choice of $\omega_0$ is critical to achieving higher representation accuracies~\cite{saragadam2023wire} and explored control over the parameters of an extended sinusoidal function for better representation quality~\cite{kazerouni2024incode}.

\subsection{SeCo-INR}

Our proposed SeCo-INR framework contains three major components: \textit{Adaptive SIREN Network}, \textit{Pixel Class Representation Network}, and \textit{Conditioner Network} which we discuss in detail below. The overall framework is demonstrated in Fig. \ref{fig_framework}.

\subsubsection{Adaptive SIREN Network}

This is the central part of the framework that maps the input coordinates to their corresponding output intensities of the image. We define this network, $f_{\theta}:\mathbb{R}^{M} \rightarrow \mathbb{R}^{P}$ as an MLP and periodic nonlinear activation functions with additional parametrization as below:
\begin{equation}
y_l=p\sin(q\omega_0(W_l y_{l-1} + b_l)+r)+s\;.
\end{equation}

Note that the above function is reduced to SIREN~\cite{sitzmann2020implicit} if $p=1, q=1, r=0, s=0$. In our framework, the values of $p,q,r,s$ are dynamically learned in every iteration of the training process, with assistance from the semantic prior information of the underlying image. Each parameter uniquely influences the characterization of the representation. The parameter $p$ sets the amplitude of the periodic signal and affects the strength of the activation, aiding in noise suppression. The parameter $q$ controls the frequency scaling of the periodic signal, focusing on the fine details of the representation and reducing high-frequency noise. The parameter $r$ adjusts the phase shift of the periodic signal, influencing the spatial arrangement in the representation. Lastly, the parameter $s$ establishes the vertical shift of the periodic signal, managing the overall contrast~\cite{kazerouni2024incode}.

\subsubsection{Pixel Class Representation Network}

In our work, we introduce a Pixel Class Representation Network to learn a continuous representation of the semantic segmentation map of a medical image. Such a segmentation map contains pixel-wise physical localization pertinent to the target image. The Pixel Class Representation Network $g(\phi)$, which is an INR model with a softmax output layer is trained to iteratively learn a continuous representation of segmentation mask $S$. It takes the coordinates $v$ as the input to generate an output semantic class distribution of those coordinates, $\hat{S}(v)=g_{\phi}(v)$. We train the weights of this network using a cross-entropy loss as $- \sum_i S(v)_i \log(\hat{S}(v)_i)$, considering $S(v)$ as the ground truth semantic class distribution for each coordinate $v$. The details about the ground truth segmentation masks are given in Section \ref{sec_datasets}.

\subsubsection{Conditioner Network}

This network $h_{\psi}:\mathbb{R}^{h}\rightarrow \mathbb{R}^{4L}$ utilizes an MLP structure and takes the local prior latent code $\hat{S}(v)\in\mathbb{R}^h$  (derived by the Pixel Class Representation Network) as input and produces the optimal parameters for each semantic region of the underlying image, \ie,  $p(v), q(v), r(v), s(v)=h_\psi(\hat{S}(v))$ which is then utilized to condition each layer of the Adaptive SIREN Network separately. The Conditioner Network learns to generate the optimum parameter values adaptively from the localized prior information given by the learned semantic segmentation map, $\hat{S}$.

\subsubsection{Loss Function}
\label{sec:loss_func}

The Loss function for the training of the overall framework is based on the reconstruction loss which enforces the Adaptive SIREN Network to fit the target image and the classification loss which enforces the Pixel Class Representation Network to learn the semantic class distributions of the coordinates. We further add a regularization term to the loss function, which mandates that the parameters $p,q,r,s$ maintain non-negative values. This type of regularization has been proven to steer the model toward more applicable solutions, accelerate model convergence, and diminish the chance of the model getting stuck in local minima during training. All trainable networks of the framework are trained in a data-driven end-to-end approach for a given target image, $I(v)$, and the overall loss function can be presented below:

\begin{gather}
\underset{\theta, \psi, \phi}{\operatorname{argmin}} \underset{v \in V}{\mathbb{E}} \left[ \| \hat{I}(v) - I(v) \|_2^2  + \beta \sum_i S(v)_i \log(\hat{S}(v)_i)\right]  \\
\text{s.t.} \quad p(v)\geq 0, \quad q(v)\geq 0, \quad r(v)\geq 0, \quad s(v)\geq 0,  ~~\forall v\notag
\end{gather}
where $\hat{I}(v)=f_\theta(v|p(v),q(v),r(v),s(v))$, $p(v), q(v), r(v), s(v)=h_\psi(\hat{S(v)})$ and $\hat{S}(v)=g_{\phi}(v)$. The hyperparameter $\beta$ determines the trade-off between reconstruction loss and the classification loss.

\subsubsection{Super-resolution Inference}

Having trained the framework and obtained the optimum model parameters, $h_{\psi}^*$, $f_{\theta}^*$, and $g_{\phi}^*$ on a given low-resolution image until convergence, the super-resolution inference can be obtained by first generating a set of extended pixel coordinates (\ie super-resolution),  $v'$. Next, the Pixel Class Representation Network can be used to generate a high-resolution semantic segmentation mask using:

\begin{equation}
\hat{S}(v')=g_{\phi}^*(v')
\end{equation}

Then, the Conditioner Network can be used to generate the optimum $p,q,r,s$ parameters for each layer of the Adaptive SIREN Network using:

\begin{equation}
%%p^*(v),q^*(v),r^*(v),s^*(v)=h_{\psi^*}(S^*(v))
p(v'),q(v'),r(v'),s(v')=h_{\psi}^*(\hat{S}(v')).
\end{equation}

Finally, the image intensities of the super-resolution image can be derived using the optimized Adaptive SIREN Network conditioned on the optimized parameters:

\begin{equation}
%%I^*(v)=f_{\theta^*}(v^*|p^*(v),q^*(v),r^*(v),s^*(v))
I(v')=f_{\theta}^*(v'|p(v'),q(v'),r(v'),s(v')).
\end{equation}

The final image, $I'$ can be formed by ordering the pixel intensities $I(v')$. The detailed algorithm of our framework is given in Algorithm \ref{alg:INR}.

% Algorithm pseudocode here
\begin{algorithm*}
\caption{SeCo-INR}
\label{alg:INR}
\begin{algorithmic}[1]
\State \textbf{Input:} low-resolution image $I$, Semantic Segmentation map $S$, Coordinates $v\in V$, hyperparameter $\beta$
\State \textbf{Output:} high-resolution image $I'$
\State Initialize the Adaptive SIREN Network $f_\theta$, Conditioner Network $h_\psi$, Pixel Class Representation Network $g_\phi$
\For{each epoch}
    \State Generate semantic segmentation class distributions: $\hat{S}(v)=g_\phi(v)$
    \State Generate $p,q,r,s$ parameters: $p(v),q(v),r(v),s(v)=h_\psi(\hat{S}(v))$
    \State Generate pixel intensities: $\hat{I}(v)=f_\theta(v|p(v),q(v),r(v),s(v))$
    \State Compute loss $\mathcal{L} = \underset{v \in V}{\mathbb{E}} \left[ \| \hat{I}(v) - I(v) \|_2^2  + \beta \sum_i S(v)_i \log(\hat{S}(v)_i)\right]$
    \State Update $\theta, \psi, \phi$ to minimize $\mathcal{L}$
\EndFor
\State Generate extended coordinates for super-resolution: $v'$
\State Generate optimized semantic segmentation class distributions: $\hat{S}(v')=g_{\phi}^*(v')$
\State Generate optimized $p,q,r,s$ parameters: $p(v'),q(v'),r(v'),s(v')=h_{\psi}^*(\hat{S}(v'))$
\State Generate pixel intensities: $I(v')=f_{\theta}^*(v'|p(v'),q(v'),r(v'),s(v'))$
\State Arrange $I(v'); \: \forall v'$ to form final image, $I'$
\State \textbf{return} $I'$
\end{algorithmic}
\end{algorithm*}

\subsection{Rationale behind feeding local semantic features }

The rationale behind the utilization of semantic information is to condition the adaptive SIREN Network with the local prior knowledge, thereby learning distinct sets of $p,q,r,s$ parameters for each semantic region of the image. For instance, the segmentation masks of a brain MRI (\eg, gray and white matter divisions, brain tumor segmentation, abnormalities, sub-cortical structures derived from a brain atlas, \etc) play a pivotal role in distinguishing different cortical regions of a brain image that exhibit various contrasts due to proton density, biophysical relaxation parameters, and geometrically varied anatomical structures~\cite{mai2015atlas,evans2012brain}. Similarly, segmentation masks of an abdominal CT image, which delineate various organs (\eg, liver, kidneys, stomach, spleen, pancreas, bowel, gallbladder, \etc) are essential for accurately analyzing the anatomical structures within the abdomen that exhibit various contrasts and geometrically varied organs~\cite{marin2014state}.

\section{Experiments}

\subsection{Datasets}
\label{sec_datasets}
Our proposed framework was validated using multiple medical imaging datasets including brain MRI datasets and abdominal CT datasets. The brain MRI datasets included fluid-attenuated inversion recovery (FLAIR) images from the fastMRI dataset~\cite{muckley2021results} as well as T1-weighted (T1-w) images from the Brain Tumor Segmentation (BraTS) datasets~\cite{menze2014multimodal}. The abdominal CT datasets are from the multi-atlas labeling challenge which had been captured during portal venous contrast phase~\cite{landman2015miccai}. For the experiments in this paper, we use the axial brain slices from the fastMRI data, which have a spatial dimension of $320 \times 320$; axial brain slices from the BraTS MRI data, which have a spatial dimension of $240 \times 240$; and axial abdominal CT slices which have a spatial dimension of $512 \times 512$. To obtain the ground truth segmentation masks, $S$, for the fastMRI brain data, we employed the FSL's Brain Extraction Tool (BET) and the Automated Segmentation Tool (FAST)~\cite{smith2002fast,jenkinson2005bet2} which uses a combination of intensity-based thresholding, surface deformation, and atlas-guided techniques. For abdominal CT data, we used the multi-atlas labeled data~\cite{landman2015miccai} which had used manually labeled 13 abdominal organs verified by a radiologist on a volumetric basis. For BraTS brain MRI data, we used the labeled data~\cite{menze2014multimodal} segmented manually and approved by experienced neuro-radiologists.

\begin{figure*}[t]
\centering
\includegraphics[width=0.98\textwidth]{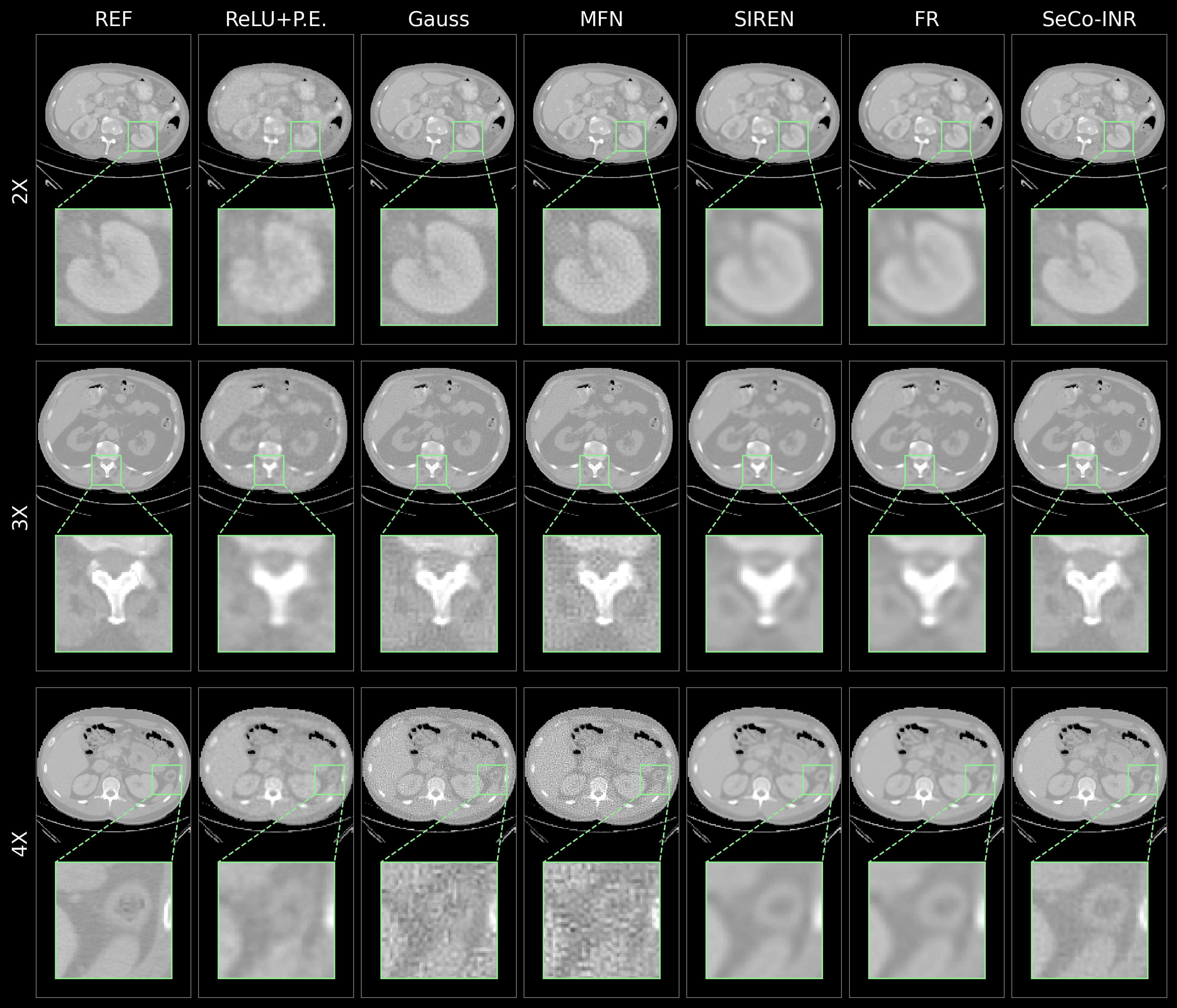}
\caption{Qualitative evaluation between different INR methods for the Abdominal CT data at resolution factors of $2\times$, $3\times$, and $4\times$. SeCo-INR produces realistic outputs of abdominal organs with sharp edges, less noise, and less blurring.}
\label{fig_abdominal}
\end{figure*}

\subsection{Experiments}

\begin{figure*}[t]
\centering
\includegraphics[width=0.98\textwidth]{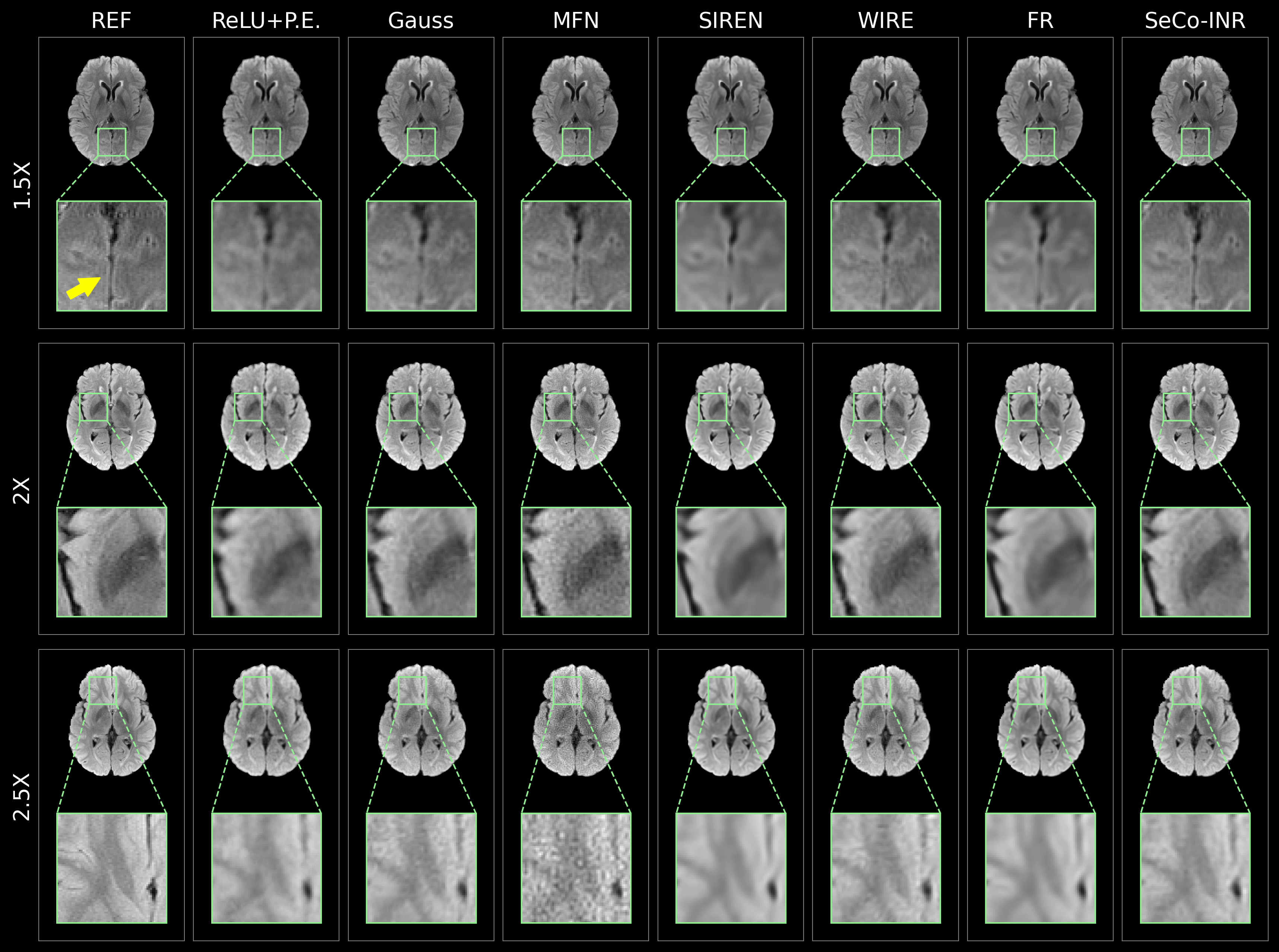}
\caption{Qualitative evaluation between different INR methods for the fastMRI brain data at resolution factors of $1.5\times$, $2\times$, and $2.5\times$. SeCo-INR produces realistic outputs of the brain with intricate anatomical details preserved (see yellow arrow), and less noise and blurring.}
\label{fig_fastmri}
\end{figure*}

\begin{figure*}[t]
\centering
\includegraphics[width=0.98\textwidth]{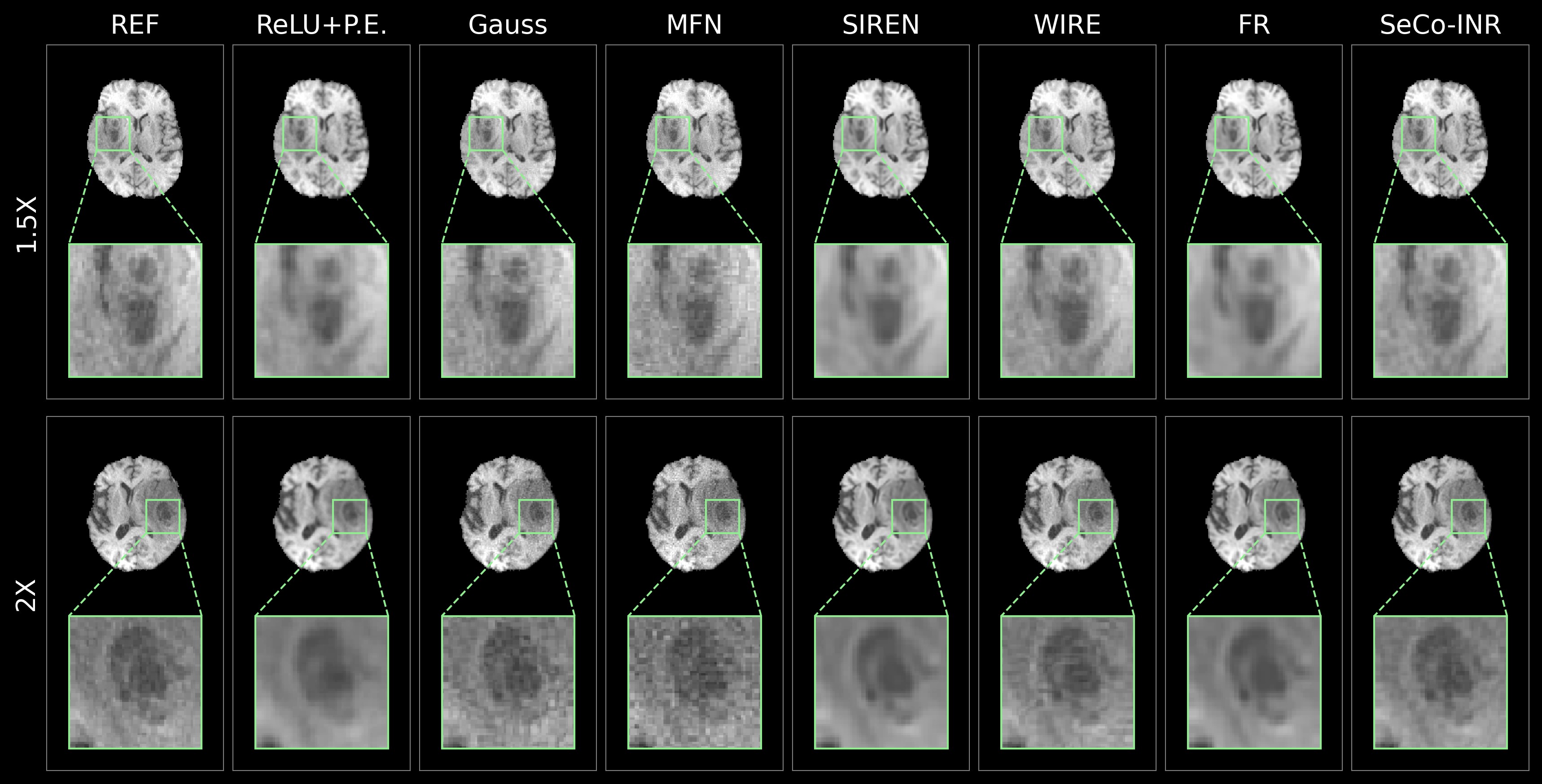}
\caption{Qualitative evaluation between different INR methods for the BraTS brain MRI data at resolution factors of $1.5\times$ and $2\times$. SeCo-INR produces realistic outputs of the tumor regions with less noise and blurring.}
\label{fig_brats}
\end{figure*}

For super-resolution experiments, we obtained images with the full spatial resolution and retrospectively resized the images by factors of $1.5\times$, $2\times$, and $2.5\times$ for fastMRI brain data, $1.5\times$ and $2\times$ for BraTS MRI brain data, and $2\times$, $3\times$, and $4\times$ for abdominal CT data. These retrospectively obtained low-resolution images contained pixelated artifacts at various degrees. Then, we implemented the super-resolution methods and assessed the outputs both qualitatively and quantitatively (\wrt to root-mean-square-error (RMSE), peak signal-to-noise ratio (PSNR) and structural similarity index (SSIM) following the standard definitions in~\cite{muckley2021results}). To compare the performance of our framework, we implemented several state-of-the-art INR-based algorithms: ReLU with Positional-Encoding (ReLU+P.E.)~\cite{tancik2020fourier}, Gauss~\cite{ramasinghe2022beyond}, Multiplicative Filter Networks (MFN)~\cite{fathony2020multiplicative}, SIREN~\cite{sitzmann2020implicit}, Wavelet INR (WIRE)~\cite{saragadam2023wire}, and Fourier Reparameterization (FR)~\cite{shi2024improved}. For the abdominal CT data, fastMRI brain data, and BraTS brain tumor data, the qualitative results are illustrated in Fig. \ref{fig_abdominal}, Fig. \ref{fig_fastmri}, and Fig. \ref{fig_brats}, respectively, and the quantitative scores are tabulated in Table \ref{tab_abdominal}, Table \ref{tab_fastmri}, and Table \ref{tab_brats}, respectively.

\begin{table*}
\centering
{\small{
\begin{tabular}{c c c c c c c c}
\hline
resolution & quantitative & \multicolumn{5}{c}{Method} \\
\cline{3-8}
factor & metric & ReLU+P.E~\cite{tancik2020fourier}  & Gauss~\cite{ramasinghe2022beyond}  & MFN~\cite{fathony2020multiplicative} & SIREN~\cite{sitzmann2020implicit}  & FR~\cite{shi2024improved} & SeCo-INR (Ours)\\
\hline
& RMSE ($\downarrow$) & 0.0207 & \underline{0.0121} & 0.0152 &	0.0153  & 0.0147 & \textbf{0.0099} \\
$2\times$ & SSIM ($\uparrow$) & 0.9335 & 0.9564 & 0.9292 & 0.9559  & \underline{0.9571} & \textbf{0.9836} \\
& PSNR ($\uparrow$) & 33.8140 & \underline{38.4625} & 36.5725 & 36.4029 & 36.7377 & \textbf{40.2139} \\
\hline
& RMSE ($\downarrow$) & 0.0233 & 0.0199 & 0.0219 & 0.0183 &  \underline{0.0181} & \textbf{0.0160} \\
$3\times$ & SSIM ($\uparrow$) & 0.9312 & 0.9018 & 0.8778 & 0.9510  & \underline{0.9518} & \textbf{0.9660} \\
& PSNR ($\uparrow$) & 32.7672 & 34.1146 & 33.3290 & 34.8399 & \underline{34.9703} & \textbf{36.0422} \\
\hline
& RMSE ($\downarrow$) & 0.0249 & 0.0306 & 0.0362 & 0.0217 & \underline{0.0216} & \textbf{0.0211} \\
$4\times$ & SSIM ($\uparrow$) & 0.9257 & 0.8046 & 0.7588 & \underline{0.9439} & \textbf{0.9445} & 0.9340 \\
& PSNR ($\uparrow$) & 32.1903 & 30.3293 & 28.9740 & 33.3819  & \underline{33.4216} & \textbf{33.6130} \\
\hline
\hline
\end{tabular}
}}
\caption{Quantitative evaluation between different INR methods for the Abdominal CT data at resolution factors of $2\times$, $3\times$, and $4\times$. (average metrics across the entire dataset are presented). The best performance is in \textbf{boldface} and the second best is \underline{underlined}.}
\label{tab_abdominal}
\end{table*}

\begin{table*}

\centering
{\small{
\begin{tabular}{c c c c c c c c c}
\hline
resolution & quantitative & \multicolumn{5}{c}{Method} \\
\cline{3-9}
factor & metric & ReLU+P.E~\cite{tancik2020fourier}  & Gauss~\cite{ramasinghe2022beyond}  & MFN~\cite{fathony2020multiplicative} & SIREN~\cite{sitzmann2020implicit} & WIRE~\cite{saragadam2023wire} & FR~\cite{shi2024improved} & SeCo-INR (Ours)\\
\hline
& RMSE ($\downarrow$) & 0.0255 & 0.0224 & 0.0217 &	0.0206 & 0.0217 & \underline{0.0202} & \textbf{0.0154} \\
$1.5\times$ & SSIM ($\uparrow$) & 0.9004 & 0.9146 & 0.9196 & 0.9195 & 0.9202 & \underline{0.9219} & \textbf{0.9606} \\
& PSNR ($\uparrow$) & 32.0217 & 33.1032 & 33.3939 & 33.8255 & 33.3758 & \underline{34.0221} & \textbf{36.3731} \\
\hline
& RMSE ($\downarrow$) & 0.0270 & 0.0258 & 0.0267 & 0.0226 & 0.0239 & \underline{0.0222} & \textbf{0.0210} \\
$2\times$ & SSIM ($\uparrow$) & 0.8951 & 0.8982 & 0.8893 & 0.9163 & 0.9117 & \underline{0.9180} & \textbf{0.9340} \\
& PSNR ($\uparrow$) & 31.5009 & 31.8457 & 31.5917 & 33.0093 & 32.5529 & \underline{33.1658} & \textbf{33.6511} \\
\hline
& RMSE ($\downarrow$) & 0.0310 & 0.0334 & 0.0364 & 0.0270 & 0.0282 & \textbf{0.0259} & \underline{0.0263} \\
$2.5\times$ & SSIM ($\uparrow$) & 0.8823 & 0.8565 & 0.8295 & 0.9024 & 0.8861 & \underline{0.9030} & \textbf{0.9030} \\
& PSNR ($\uparrow$) & 30.3102 & 29.5691 & 28.8997 & 31.4860 & 31.0746 & \textbf{31.8485} & \underline{31.6760} \\
\hline
\hline
\end{tabular}
}}
\caption{Quantitative Evaluation between different INR methods for the fastMRI brain data at resolution factors of $1.5\times$, $2\times$, and $2.5\times$. (average metrics across the entire dataset are presented). The best performance is in \textbf{boldface} and the second best is \underline{underlined}.}
\label{tab_fastmri}
\end{table*}

\begin{table*}
\centering
{\small{
\begin{tabular}{c c c c c c c c c}
\hline
resolution & quantitative & \multicolumn{5}{c}{Method} \\
\cline{3-9}
factor & metric & ReLU+P.E~\cite{tancik2020fourier}  & Gauss~\cite{ramasinghe2022beyond}  & MFN~\cite{fathony2020multiplicative} & SIREN~\cite{sitzmann2020implicit} & WIRE~\cite{saragadam2023wire} & FR~\cite{shi2024improved} & SeCo-INR (Ours)\\
\hline
& RMSE ($\downarrow$) & 0.0248 & 0.0220 & 0.0234 &	0.0213 & \underline{0.0207} & 0.0208 & \textbf{0.0191} \\
$1.5\times$ & SSIM ($\uparrow$) & 0.9436 & 0.9556 & 0.9409 & 0.9546 & \underline{0.9626} & 0.9562 & \textbf{0.9732} \\
& PSNR ($\uparrow$) & 32.1597 & 33.1880 & 32.6338 & 33.4810 & \underline{33.7224} & 33.6702 & \textbf{34.4158} \\
\hline
& RMSE ($\downarrow$) & 0.0281 & 0.0279 & 0.0350 & 0.0205 & 0.0225 & \textbf{0.0201} & \underline{0.0205} \\
$2\times$ & SSIM ($\uparrow$) & 0.9219 & 0.9170 & 0.8797 & 0.9559 & 0.9470 & \underline{0.9577} & \textbf{0.9604} \\
& PSNR ($\uparrow$) & 31.0791 & 31.0991 & 29.1585 & 33.7836 & 32.9758 & \textbf{33.9721} & \underline{33.7852} \\
\hline
\hline
\end{tabular}
}}
\caption{Quantitative evaluation between different INR methods for the BraTS brain MRI data at resolution factors of $1.5\times$ and $2\times$. (average metrics across the entire dataset are presented). The best performance is in \textbf{boldface} and the second best is \underline{underlined}.}
\label{tab_brats}
\end{table*}

\subsection{Implementation}

In our experiments, a 5-layer MLP was utilized as the Adaptive SIREN Network consisting of 256 neurons in each layer. To implement the constraints on the $p,q,r,s$ parameters, we enforced ReLU functions and added to the overall loss to penalize the generation of negative values~\cite{kazerouni2024incode}. We implemented the framework on a single NVIDIA GeForce RTX 3060 GPU with 12GB memory using the PyTorch framework. The optimization process involved using the Adam optimizer with a learning rate scheduler, \ie decreasing the learning rate by 0.1 after each epoch. We set $\beta=1$. For convergence, ReLU+P.E., Gauss, and MFN models were trained for 2000 epochs, whereas SIREN, WIRE, FR, and SeCo-INR models were trained for 1000 epochs. All codes for reproducibility are attached to the Supplementary Materials.

\subsection{Ablation Studies}

We conducted an ablation study to demonstrate that the performance gains are indeed due to the incorporation of semantic segmentation information in the INR framework. We utilized the extended SIREN network without feeding semantic features into the model. Table \ref{tab_ablate} presents the quantitative scores for the fastMRI data at a resolution factor of $2\times$ and they clearly indicate that the performance gains stem from the semantic features. This validates that a key strength of SeCo-INR is its ability to leverage additional strong medical image segmentation information.

We also compared the computational costs of each algorithm on the fastMRI brain dataset at a resolution factor of $2\times$ by calculating the time taken on average, per image to reach a PSNR level of $30.0$ OR an SSIM level of $0.90$. The results are presented in Table \ref{tab_comptime}.

\begin{figure}
\centering
\includegraphics[width=0.45\textwidth]{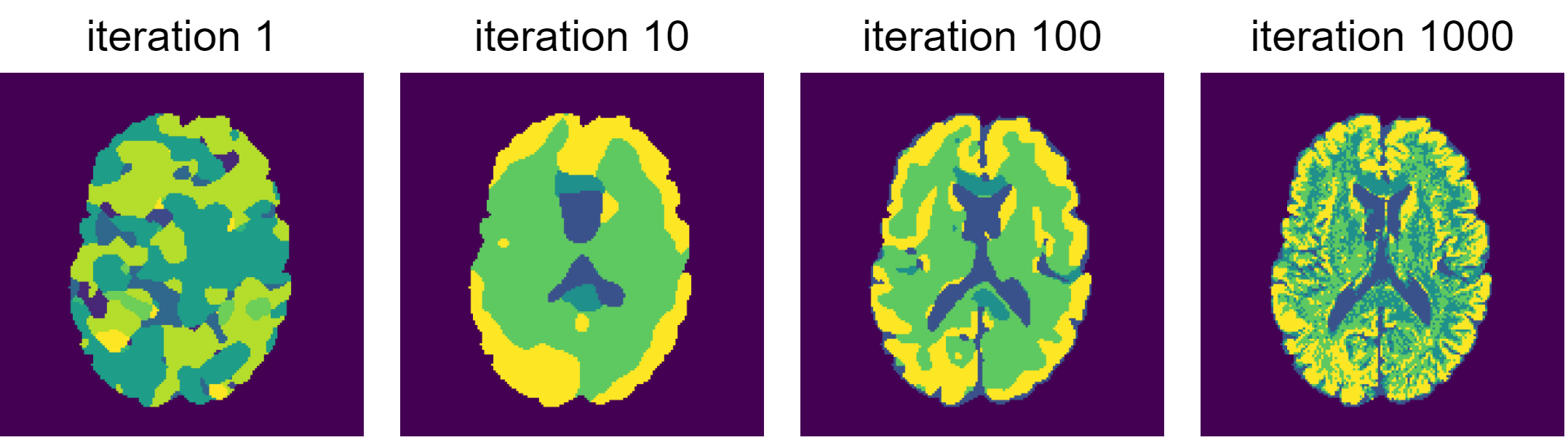}
\caption{An example from the fastMRI brain dataset shows the evolution of the segmentation mask learning process using the Pixel Class Representation Network.}
\label{fig_segmasks}
\end{figure}

Our overall framework simultaneously learns the semantic segmentation mask as well as the underlying image using the Pixel Class Representation Network and the Adaptive SIREN Network, respectively. Figure \ref{fig_segmasks} demonstrates an example segmentation mask from the fastMRI brain dataset learned iteratively.

\begin{table*}
\centering
{\small{
\begin{tabular}{c c c c c c c}
\hline
 ReLU+P.E~\cite{tancik2020fourier}  & Gauss~\cite{ramasinghe2022beyond}  & MFN~\cite{fathony2020multiplicative} & SIREN~\cite{sitzmann2020implicit} & WIRE~\cite{saragadam2023wire} & FR~\cite{shi2024improved} & SeCo-INR (Ours)\\
\hline
 4.19 & \underline{0.51} &	3.39 & 0.65 & 4.66 & 0.66 & \textbf{0.45} \\
						
\hline
\hline
\end{tabular}
}}
\caption{Comparison of computational time (in seconds) of each algorithm for $2\times$ super-resolution on the fastMRI brain data. The best performance is in \textbf{boldface} and the second best is \underline{underlined}.}
\label{tab_comptime}
\end{table*}

\begin{table}
  \centering
  {\small{
  \begin{tabular}{c c c c}
    \hline
     & RMSE ($\downarrow$) & SSIM ($\uparrow$) & PSNR ($\uparrow$) \\
    \hline
    w/o semantic info & 0.0231 & 0.9112 & 32.8125\\
    SeCo-INR & 0.0210 & 0.9340 & 33.6511 \\
    \hline
    \hline
  \end{tabular}
  }}
  \caption{Ablation study: Comparison of quantitative scores on the fastMRI data at a resolution factor of $2\times$ with and without feeding semantic information to the extended INR model.}
  \label{tab_ablate}
\end{table}

\section{Results and Discussion}

As seen in Fig. \ref{fig_abdominal}, SeCo-INR is able to achieve high-quality CT super-resolution with minimal noise and blurring. The $2\times$ example highlights the kidney region where Seco-INR is able to reconstruct the edges and contrast information accurately, whereas Gauss and MFN methods produce noisy outputs, and SIREN and FR methods produce blurry outputs. The $3\times$ example highlights the spinal region where most of the pixelated artifacts are rectified in the SeCo-INR output compared to other methods. The $4\times$ example highlights the spleen region where SeCo-INR demonstrates superiority in mitigating noise during super-resolution and producing realistic outputs. These observations align with the quantitative scores in Table \ref{tab_abdominal}, where SeCo-INR holds the best performance in almost all super-resolution levels and metrics. 

As depicted in Fig. \ref{fig_fastmri} and Fig. \ref{fig_brats}, our framework demonstrates the overall best image resolution and SNR for the super-resolution task in MRI, especially at higher resolution levels. While all methods successfully rectify the pixelated artifacts of the low-resolution MRI images, MFN, Gauss, WIRE, and FR methods produce noisier images and ReLU+P.E. and SIREN methods tend to overly smooth the image, resulting in failure to reconstruct intricate anatomical structures of the brain (see yellow arrow in the $2\times$ example). In contrast, SeCo-INR accurately reconstructs intricate edge details and eliminates pixelation artifacts, enabling higher-resolution MRI images with minimal noise. These results are further confirmed by the quantitative scores presented in Table \ref{tab_fastmri}, where SeCo-INR outperformed all other methods at  $1.5\times$ and $2\times$ super-resolution, \wrt RMSE, PSNR, and SSIM. The brain tumor images in Fig. \ref{fig_brats} show similar observations where SeCo-INR demonstrates the overall best image resolution and SNR specifically in the tumor region of the brain. Methods like ReLU+P.E., SIREN, and FR methods tend to overly smooth the tumor region whereas the other methods produced noisy outputs hindering the accurate visibility of the brain tumor, especially at $2\times$ super-resolution. The quantitative scores confirmed these observations where SeCo-INR dominated in all the metrics computed.

As evident by the results in Table \ref{tab_comptime}, our SeCo-INR framework has the added benefit of faster convergence where it reaches certain threshold quantitative scores using minimal training time. This confirms SeCo-INR as a viable solution in practical medical diagnostic environments where faster computation is critical. Not only in training, SeCo-INR provides the additional benefit of automatically selecting the activation hyperparameters in contrast to other methods like SIREN and WIRE which are extremely sensitive to the activation hyperparameters~\cite{saragadam2023wire} and are difficult to fine-tune for a given dataset.

One of the limitations of SeCo-INR is the requirement of ground truth semantic segmentation masks. For the experiments in this paper, we utilized the readily available segmentation masks or generated masks using existing tools. Nevertheless, many other tools are available to obtain segmentation masks for a variety of anatomical regions, including simple thresholding and clustering techniques, atlas-based methods~\cite{gubern2009atlas}, or recent AI tools such as MedSAM~\cite{ma2024segment} and MONAI~\cite{cardoso2022monai}.

\section{Conclusion}

This paper presents SeCo-INR, an INR-based medical image super-resolution method that leverages local prior information derived from a semantic segmentation mask of the underlying image. SeCo-INR enables an INR network to learn local region-specific parameters for the underlying medical image, thereby achieving accurate super-resolution. The results have been validated both qualitatively and quantitatively on MRI and CT data, and the method appears to be a promising solution for medical image super-resolution.

% \section{Acknowledgements}
% We acknowledge the support of the Australian Research Council (ARC) Fellowship Program IM230100002 as well as the ARC grants LP170100494 and DP210101863.

% \clearpage

%%%%%%%%% REFERENCES
{\small
\bibliographystyle{ieee_fullname}
\bibliography{egbib}
}

\end{document}